\documentclass[submission,copyright,creativecommons]{eptcs}
\providecommand{\event}{ACL2 2018} % Name of the event you are submitting to
\usepackage{breakurl}             % Not needed if you use pdflatex only.
\usepackage{underscore}           % Only needed if you use pdflatex.

\title{Incremental SAT Library Integration Using Abstract Stobjs}
\author{Sol Swords
\institute{Centaur Technology, Inc.}
\email{sswords@centtech.com}
}
\def\titlerunning{Incremental SAT Library Integration Using Abstract Stobjs}
\def\authorrunning{S. Swords}
\begin{document}
\maketitle

\begin{abstract}
We describe an effort to soundly use off-the-shelf incremental SAT
solvers within ACL2 by modeling the behavior of a SAT solver library
as an abstract stobj.  The interface allows ACL2 programs to use
incremental SAT solvers, and the abstract stobj model allows us to
reason about the behavior of an incremental SAT library so as to
show that algorithms implemented using it are correct, as long as
the library is bug-free.
\end{abstract}

\section{Introduction}

ACL2 users have long recognized the potential utility of integrating
external proof tools \cite{KAUFMANN20093}.  While many such tools have
been successfully integrated and used in ACL2 proofs, as far as we
know these have all been used in a stateless manner: that is, a
complete query is built up within ACL2 and exported in a format
accessible to the external tool, then the external tool is executed
and finishes, at which point ACL2 reads its output.  However, some
external tools benefit from storing state between queries.
Incremental satisfiability (SAT) solvers, in particular, keep learned
facts and heuristic information from previous queries and use these to
speed up later queries, allowing for repeated SAT checks that can be
much faster than they would be if the solver was started from scratch
for every query.

This paper describes an interface allowing ACL2 programs to use an
external incremental SAT library in a stateful manner.  The library is
accessed through an abstract stobj.  The abstract stobj interface
functions' logical definitions model the behavior of the incremental
SAT library, and their executable definitions call into the library to
set up queries and get their results.  We show how our model of an
incremental SAT solver can be used in a complete algorithm, namely
\textit{fraiging} or AIG SAT sweeping
\cite{Mishchenko:2006:ICE:1233501.1233679}, and the algorithm proved
correct assuming correct behavior of the incremental solver.

%  our integration is sound, we claim that
% it could, in principle, be sound, and t  We have taken care to block all avenues of
% potential unsoundness that we have discovered.  In Section
% \ref{section:nondeterminism} we describe our effort to model
% nondeterministic behavior of the external library, which is a
% potential soundness pitfall of any external tool.  In Section
% \ref{section:backend} we discuss the potential for unsoundness
% from overwriting the executable definitions of the abstract stobj
% interface functions to call the backend solver, and our efforts to
% prevent this.
 
We begin in Section \ref{section:ipasir} by discussing the incremental
SAT library interface we are targeting.  Next we give an overview of
the integration in Section \ref{section:overview}, describe the
logical model in Section \ref{section:logicalmodel}, and describe the
mechanics of interfacing with the external library in Section
\ref{section:backend}.  We assess the soundness of the incremental SAT
integration in Section \ref{section:soundness}.  We then describe the
implementation of a fraiging algorithm using incremental SAT in
Section \ref{section:fraiging}.

% we have explored
% several avenues for potential unsoundness and  and argue that it can be considered
% to be \textit{potentially sound}; that is, if the external library
% correctly

% We have modeled an incremental SAT solver in ACL2 
% We use the example of incremental SAT to illustrate a way to build an
% interface between ACL2 and a stateful external program.  The external
% program is logically modeled as an abstract stobj, and the program can
% be queried and manipulated using the abstract stobj accessor and
% updater functions.  

% The integration of this external tool is known to be unsound due to
% potential misuse of the concrete stobj upon which the abstract stobj
% is based.  We discuss that source of unsoundness in Section
% \ref{section:unsoundness}.  But this particular soundness problem is
% easy to avoid and easy to detect -- the concrete stobj simply
% shouldn't be used except in defining the abstract stobj.  This problem
% could also potentially be solved by adding an ACL2 feature that would
% allow use of the concrete stobj to be disabled.

% A more immediate consideration is whether the intended use of the
% library can result in unsoundness.  We can't prove that it won't: at a
% minimum, we need to trust that the external library acts as our model
% assumes that it does.  We also need to carefully account for
% nondeterminism, which we discuss in Section
% \ref{section:nondeterminism}.  We 

\section{Related Work}

Several other efforts have resulted in integrations between ACL2 and
external proof tools.  Our work is most directly based on SATLINK
\cite{EPTCS114.8}, which calls an external SAT solver executable on a
single problem in a stateless manner.  SATLINK provides a function
that calls an external SAT solver on a CNF formula; that function is
assumed to only return \texttt{:unsat} when the formula is
unsatisfiable, and this can be used to perform ACL2 proofs using GL
\cite{DBLP:journals/corr/abs-1110-4676} or by otherwise appealing to that
assumption.  Similarly, SMTLINK \cite{DBLP:journals/corr/PengG15}
provides a trusted clause processor which encodes ACL2 formulas as SMT
problems and calls the Z3 SMT solver to prove them, also statelessly.
Reeber's SAT-based decision procedure for a decidable subset of ACL2
formulas \cite{sulfa} also calls an external SAT solver
statelessly.

Somewhat different in flavor is ACL2SIX \cite{4021022}, which calls
IBM's internal SixthSense checker to verify hardware properties.
SixthSense, in this case, provides not just the decision procedure but
the semantics of the model as well.  The ACL2 logical interface to the
hardware model consists of two functions \texttt{sigbit} and
\texttt{sigvec}.  These functions each take as inputs an \textit{entity} representing a
machine and environment model known to the external tool, a signal name, and a time,
and they return the value of the
signal at the given time.  These functions are not defined and cannot
be executed, but facts about them can be obtained by calling the
ACL2SIX clause processor.  This clause processor renders the formula
into a VHDL property and calls SixthSense to prove the property.  For
an adder module, for example, the \texttt{sum} signal at time $n$ can
be proven to be the bit-vector sum of the \texttt{a} and \texttt{b}
inputs at time $n-1$.  Calls into SixthSense are stateless, but
because of the simplicity of the logical connection between the
external solver and ACL2, a stateful integration could have the same
logical story.  That is, integration with a SixthSense shared
library which collected information about a hardware model across
multiple queries could be used with the same logical model.

\section{Incremental SAT Interface}
\label{section:ipasir}

Rather than targeting one particular incremental SAT solver, we chose
to interface with IPASIR, a simple C API introduced for use in SAT Race
2015 \cite{balyo2015} and used through the 2017 SAT
competition \cite{balyo2017}.  (IPASIR stands for
Reentrant Incremental SAT solver API, in reverse.) The IPASIR
interface consists of the following 10 functions. The API describes
the states of the solver object as
\texttt{INPUT}, \texttt{SAT}, or \texttt{UNSAT}.  Most functions
may be used in any of these states, but a few, as noted
below, require the solver to be in a particular state.

\begin{itemize}
\item \texttt{ipasir_signature} returns a name and version string for the solver library.

\item \texttt{ipasir_init} constructs a new solver object in the \texttt{INPUT} state and returns a pointer to it.

\item \texttt{ipasir_release} destroys a solver object.

\item \texttt{ipasir_add} adds a literal to the new clause currently being built or, if
  the input is 0 (which is not a literal), adds that clause
  to the formula; the resulting solver is in the \texttt{INPUT} state.

\item \texttt{ipasir_assume} adds a literal to be assumed true during the next SAT query 
and puts the solver in the \texttt{INPUT} state.

\item \texttt{ipasir_solve} solves the formula under the current
  assumptions, determining whether it is satisfiable or unsatisfiable
  unless it is interrupted by the \texttt{ipasir_set_terminate}
  callback.  Puts the solver into the \texttt{INPUT} state if the check failed due to a termination condition,
  \texttt{SAT} or \texttt{UNSAT} respectively if satisfiable or
  unsatisfiable.

\item \texttt{ipasir_val} returns the truth value of a variable in the
  satisfying assignment produced by the previous call of
  \texttt{ipasir_solve}; it requires that the solver is in the
  \texttt{SAT} state and leaves it in that state.

\item \texttt{ipasir_failed} checks whether a given assumption literal
  was used in proving the previous unsatisfiability result produced by
  \texttt{ipasir_solve}; it requires that the solver is in the
  \texttt{UNSAT} state and leaves it in that state.

\item \texttt{ipasir_set_terminate} sets up a callback function that
  will be called periodically by the solver during
  \texttt{ipasir_solve} and can decide whether to interrupt the
  search.  Preserves the current state of the solver.

\item \texttt{ipasir_set_learn} sets up a callback function that will
  be called each time a clause is learned, allowing these clauses to
  be recorded.  Preserves the current state of the solver.

\end{itemize}

The IPASIR interface supports the following basic usage of an incremental
solver.  The client first creates a solver object using
\texttt{ipasir_init}, then builds up a formula using repeated calls of
\texttt{ipasir_add}.  Usually the formula itself should be
satisfiable: clauses can only be added and not removed, so once it is determined that the
formula is unsatisfiable, no further information can be obtained from subsequent queries---it will always remain unsatisfiable.
Instead, the formula is kept satisfiable, but a separate set of assumptions may be provided using \texttt{ipasir_assume} that may or may not be satisfiable in conjunction with the formula.  A call to 
\texttt{ipasir_solve} checks whether the formula and all the
assumptions can be simultaneously satisfied, after which the assumptions are deleted.
When the call of
\texttt{ipasir_solve} returns satisfiable, the satisfying assignment
may be queried using \texttt{ipasir_val}, and when unsatisfiable, the
unsatisfiable subset of the assumptions may be queried using
\texttt{ipasir_failed}.  After the client is done with these queries,
it can then add more clauses and/or assumptions before solving again.
(Once the client starts adding clauses or assumptions, it is no longer
allowed to query the satisfying assignment or unsatisfiable subset 
until after the next call to \texttt{ipasir_solve}.)
When done, the client calls \texttt{ipasir_release} to free the
solver's memory.

Our ACL2 interface supports this usage pattern, but it does not yet
support \texttt{ipasir_set_learn} and only supports
\texttt{ipasir_set_terminate} in a limited fashion, allowing searches
to be interrupted after the callback is called some number of times.
Some of the other functions are modified so as to make their
output more idiomatic in ACL2.  For example, the C interface for
\texttt{ipasir_solve} returns 10 when satisfiable, 20 when
unsatisfiable, and 0 when interrupted; we instead return one of the
symbols \texttt{:sat}, \texttt{:unsat}, or \texttt{:failed}.

\section{Overview of ACL2 Integration}
\label{section:overview}

The logical model of the IPASIR interface is contained in a book named
\texttt{ipasir-logic.lisp}, separate from the code which interfaces
with the external library.  This book is purely in the ACL2 logic,
and its only trust tag is used to flag a function as having an unknown
constraint, which in itself cannot cause unsoundness.
Books defining programs that use IPASIR
functionality can therefore remain free of external
code or under-the-hood hacks.
  They will simply require that the book defining the executable
backend interface, \texttt{ipasir-backend.lisp}, be loaded before
any IPASIR functions can be actually executed.

The \texttt{ipasir-logic} book defines an abstract stobj that provides
the interface to the incremental SAT library.  Abstract stobjs were
conceived as a mechanism to replace a complicated executable
implementation with a simpler logical model that still accurately
represents the functionality of that implementation \cite{EPTCS114.5}.
Our use of abstract stobjs for interfacing with an external library is
in this same spirit, differing in that the executable definitions used in admitting the abstract stobj
are later replaced (when \texttt{ipasir-backend} is loaded) by under-the-hood
implementations that use the external library.
%  but when the backend code is loaded, it uses an
% under-the-hood executable implementation for each of the functions in
% the abstract stobj interface instead of one in the ACL2 logic.

The features of abstract stobjs are a very good fit for the requirements of
interfacing with an external library:
\begin{itemize}
\item Usage of the library is forced to follow the stobj discipline,
  i.e., any operation that changes the state of the external solver
  object must return that object, and that return value must replace
  the (single) existing binding of that object.
\item The set of interface functions can be restricted to a small API.
  This allows the logical model to contain data that can't be
  accessed in executable code (and therefore need not be computed) but
  may still be important for modeling the behavior.
\item The interface functions' guards may be engineered to prevent
  ill-defined behaviors and illegal interactions.
\end{itemize}

% Using an abstract stobj as the model for such an interface offers many
% benefits:

% \item It offers a convenient encapsulation of the functions that need
%   to be replaced with under-the-hood definitions -- namely, the
%   abstract stobj \texttt{:exec} functions.

% t
% allows the interface to be restricted to only a few supported
% functions.  That way, the logical model may contain data that is not
% accessible (and therefore does not need to be computed) but helps to
% enforce sound behavior.

%   The logical model of the abstract stobj is simply a cons
% object satisfying a certain product type predicate
% \texttt{ipasir\$a-p}.  Many of the fields of that product will
% not actually be accessible via the abstract stobj interface but are
% still important to correctly modeling the behavior of the library.

% under-the-hood mechanism which loads a solver library and overwrites
% the executable code of each interface function with a function that
% calls into the solver library.  We describe the logical model in
% Section \ref{logical_model_section} and briefly describe
% the interfacing mechanism in Section \ref{library_interface_section}.  Since the logical model is in 

\section{Logical Model}
\label{section:logicalmodel}
Our logical model of an incremental SAT solver is built on the
existing theory of conjunctive normal form satisfiability provided by
the SATLINK library \cite{EPTCS114.8}.  In particular, a literal is
represented as a natural number with its least significant bit
representing its polarity and the rest of its bits giving the index of its
variable; a clause is a list of literals, and a formula is a list of
clauses.  The representation of literals is a trivial difference
between the ACL2 interface and the C interface, which uses signed
non-zero integers with negation giving the sign and the absolute value
giving the variable number.  We will refer to these C-style literals as DIMACS
literals, as opposed to SATLINK literals.

The IPASIR solver is accessed via an abstract stobj called
\texttt{ipasir}.  The logical model of that abstract stobj is an
object of type \texttt{ipasir\$a}, which is a product type containing
the following fields:

\begin{itemize}
\item \texttt{formula}, a list of lists of literals, representing the
  permanent formula stored in the solver

\item \texttt{assumption}, a list of literals, the current assumption

\item \texttt{new-clause}, a list of literals, the current clause being built

\item \texttt{status}, one of the symbols \texttt{:undef},
  \texttt{:input}, \texttt{:unsat}, or \texttt{:sat}, the current
  state of the solver

\item \texttt{solution}, a list of literals, either a satisfying
  assignment or an unsatisfiable subset of the previous assumption

\item \texttt{solved-assumption}, a list of literals, the previous
  assumption that was proved unsatisfiable (used in the guard for
  \texttt{ipasir-failed}, described below)

\item \texttt{callback-count}, the number of times the callback function
  (set in the underlying solver with \texttt{ipasir_set_terminate})
  has been called during solve

\item \texttt{history}, a list representing the history of all
  operations performed on the solver object.
\end{itemize}

The abstract stobj interface only allows direct access to a few of
these fields.  The others are logical fictions which are
convenient for modeling the state of the underlying solver
implementation but are never actually built.  We describe the full
abstract stobj interface below, but we begin by briefly describing how
the history field is updated.

The history field is a record of all updates performed on the
solver object.  Each updater operation adds an entry to the history,
which ensures that the solver object is never equal to a previous
version of itself.  Additionally, before the solver object can be used
the history must be initialized with an object read from the ACL2
state's oracle, which prevents any two solvers from being provably
equal.  This removes a source of unsoundness due to nondeterminism,
discussed in more detail in Section \ref{section:nondeterminism}.  In
our description of the operations below, we will omit discussion of
the history field because it is always updated by simply consing on a
new entry.

%   As with all stobjs, the event that defined the
% \texttt{ipasir} stobj also creates an instance of that stobj.
% However, all of the interface functions for the abstract stobj have
% guards that can neither be executed nor proved to hold of that global
% stobj instance, so that instance can't be used without doing something
% illegal.  (This is a side effect of restrictions used to ensure that
% the library is used in a well-defined manner.)  Instead, local
% \texttt{ipasir} objects may be created within a function using
% \texttt{with-local-stobj}.  This creates a new stobj using
% \texttt{create-ipasir}.

%   The new stobj created is well-defined, with
% all fields having their default values: \texttt{status} is
% \texttt{:undef}, \texttt{callback-count} is 0, and all other fields
% are \texttt{NIL}. 
% There is only one interface function whose guards
% allow it to be called on an object in this state: \texttt{(ipasir-init
%   ipasir state)}, whose guard only requires that the solver's status
% is \texttt{:undef}.
We now describe how a solver object is initialized and released.  When
it is created, an \texttt{ipasir} stobj initially has status
\texttt{:undef} and an empty history.  The only interface function
that allows us to progress from this state is \texttt{ipasir-init},
whose guard only requires that the status be \texttt{:undef}.
\texttt{ipasir-init} initializes all non-history fields to default
values except it sets status to \texttt{:input}.  It uses the ACL2
state to add to the history a value read from the state's oracle
field, and returns a modified state with that value removed from the
oracle.  The solver object is then usable, having status
\texttt{:input}.  When done, it should be freed using
\texttt{ipasir-release}, which sets the status field back to
\texttt{:undef}.  (If a \texttt{with-local-stobj} form creating an
\texttt{ipasir} object is exited without calling
\texttt{ipasir-release}, the memory used by the backend solver object
will be
leaked.)  The solver can be reinitialized after releasing it using
\texttt{ipasir-reinit}, which is much like \texttt{ipasir-init}, but
does not take state and cannot be used for the first initialization
(it requires that the history field is non-empty).

The remaining functions support the basic usage model of an
incremental solver.  The following functions are used to set up the
problem to be solved; they may be used in any initialized state (i.e.,
their guards only require that status is not \texttt{:undef}), and all
set the status to \texttt{:input}:

\begin{itemize}
\item \texttt{(ipasir-add-lit ipasir lit)} conses the given literal
  onto the \texttt{new-clause} field.
\item \texttt{(ipasir-finalize-clause ipasir)} adds the current
  \texttt{new-clause} to the formula and empties it.
\item \texttt{(ipasir-assume ipasir lit)} conses the given literal
  onto the \texttt{assumption} field.
\item \texttt{(ipasir-input ipasir)} only sets the status to :input.
\end{itemize}

The \texttt{ipasir-input} is convenient when defining functions that
may add some assumptions or clauses but sometimes do nothing; in this
case, if one calls \texttt{ipasir-input} instead of doing nothing,
then the status of the resulting solver will always be
\texttt{:input}.  This is allowable since any interface function that
can be called in the \texttt{:input} state may be called in any
state other than \texttt{:undef}.

After setting up the problem, \texttt{(ipasir-solve ipasir)} is used
to check satisfiability.  This is a constrained function which returns
a search status as well as a new solver object.  Its guard requires
that status is not \texttt{:undef} and that the \texttt{new-clause} is
empty.  (This requirement simply removes an ambiguity from the
interface specification: it isn't clear what it means if some literals
have been added to a new clause but the clause has not been finalized when the
solver is called.)  The constraints require:
\begin{itemize}
\item The search status returned will be \texttt{:failed},
  \texttt{:unsat}, or \texttt{:sat}.
\item The resulting solver will have status \texttt{:input} if failed,
  \texttt{:unsat} or \texttt{:sat} correspondingly otherwise.
\item If \texttt{:unsat}, the solution field of the result solver
  contains a subset of the input solver's assumption, and that subset
  cannot be satisfied in conjunction with the formula.
\item If \texttt{:unsat}, the solved-assumption field of the new
  solver equals the assumption field of the input solver.
\item The assumption and new-clause fields of the resulting solver are
  empty, and the formula is preserved from the input solver.
\item The callback count of the resulting solver is greater than or
  equal to that of the input solver.
\end{itemize}

We could assume that the solver produces a satisfying assignment when
it returns \texttt{:sat}, but in most applications it is easy to check
that the assignment is correct, if necessary.

The constraints of the \texttt{ipasir-solve} function do not fully
determine its behavior---for example, it is allowed to return
\texttt{:failed} on any input, even when it could alternatively return
\texttt{:sat} or \texttt{:unsat}.  When actually executed with a
backend solver loaded, \texttt{ipasir-solve} will return the answer
supplied by the solver---this gives us access to facts about
\texttt{ipasir-solve} that are not implied by its constraints.  In
ACL2's usual treatment of constrained functions, the constraints are
assumed to be everything that is known about the function \textit{a
  priori}, and to imply everything that is later proved about it.  Any
facts proved about a constrained function can then be
\textit{functionally instantiated}, i.e., assumed true of any other
function that satisfies the constraints \cite{boyer1991}.
Since this is not true of the constraints for \texttt{ipasir-solve},
 we say that it has \textit{unknown constraints}\footnote{
  The unknown constraint is currently added using the
  \texttt{define-trusted-clause-processor} utility, but it may be
  supported more directly by the ACL2 system in the future.}
\cite{KAUFMANN20093}, which prevents functional instantiation of
facts we have proved about it.

After solving, if the result was \texttt{:sat} or \texttt{:unsat}, the
solver may be queried to derive the satisfying assignment or the
unsatisfiable subset of the assumptions, respectively:

\begin{itemize}
\item \texttt{(ipasir-val ipasir lit)} requires that status is
  \texttt{:sat} and returns 1, 0, or \texttt{NIL} depending whether
  that literal is true, false, or undefined in the satisfying
  assignment (i.e. the solution field).
\item \texttt{(ipasir-failed ipasir lit)} requires that status is
  \texttt{:unsat} and that the literal is a member of the solver's
  solved-assumption field, and returns 1 if the literal is a member of
  the identified unsatisfiable subset (i.e. the solution field), 0 if
  not.
\end{itemize}

In many algorithms it's desirable to limit the amount of time spent
trying to solve any one query.  We support this via the function
\texttt{(ipasir-set-limit ipasir limit)}, where limit is a natural
number or \texttt{NIL}; passing a natural number here will cause the
solver to fail a call of \texttt{ipasir-solve} after that many
callbacks, and passing \texttt{NIL} will remove that limit.
Logically, this only affects the history and resets the callback count
to 0.  The callback count can be accessed for performance monitoring
using \texttt{(ipasir-callback-count ipasir)}.

Rob Sumners contributed an update to
the IPASIR integration that adds to the abstract stobj interface the
functions necessary to make all the guards executable.
In earlier versions, the guards for the
interface functions were all non-executable, which in practice meant
that all execution must be done on \texttt{ipasir} objects created by
\texttt{with-local-stobj}, not the global \texttt{ipasir} object.
  An \texttt{ipasir} object newly created by
\texttt{with-local-stobj} is known to be in a certain state, so
functions that use this mechanism could have verified, executable guards
even if they called ipasir interface functions that have
non-executable guards.  For example, the guard for \texttt{ipasir-init} is:
\begin{verbatim}
  (non-exec (eq (ipasir$a->status ipasir) :undef))
\end{verbatim}%$
This needed to be non-executable because \texttt{ipasir\$a->status} is just the logical model, which can't be executed on the stobj, and there was no abstract stobj interface function
that could return the status or check whether it was \texttt{:undef}.  However, a function 
with verified, executable guards could be created using \texttt{with-local-stobj} as follows, 
because the \texttt{ipasir} object created is known to have \texttt{:undef} status:
\begin{verbatim}
(defun ipasir-initialize-and-release (state)
  (declare (xargs :stobjs state))
  (with-local-stobj ipasir
    (mv-let (ipasir state ans)
      (b* (((mv ipasir state)
            (ipasir-init ipasir state))
           (ipasir (ipasir-release ipasir)))
        (mv ipasir state nil))
      (mv state ans))))
\end{verbatim}

Sumners added the following interface functions, which suffice to express all the guards as executable terms:

\begin{itemize}
\item \texttt{ipasir-get-status} returns the solver status, \texttt{:undef}, \texttt{:input}, \texttt{:unsat}, or \texttt{:sat}
\item \texttt{ipasir-some-history} returns \texttt{T} if the history is nonempty
\item \texttt{ipasir-empty-new-clause} returns \texttt{T} if the \texttt{new-clause} is empty
\item \texttt{ipasir-get-assumption} returns the list of current assumption literals
\item \texttt{ipasir-solved-assumption} returns the assumption before the last solve, if the solver produced \texttt{:unsat}.
\end{itemize}

Sumners additionally added two extra interface functions that require
library support that is not part of the IPASIR API, but still can be
easily supported by most incremental SAT libraries.  If the external
library is set up to support these, then an extra book can be loaded
which supplies their actual implementation; otherwise, stub functions
are used instead.  The two functions:

\begin{itemize}
\item \texttt{ipasir-bump-activity-vars} increases the activity
  heuristic of the variables of the given literals
\item \texttt{ipasir-get-curr-stats} returns several counters useful
  for heuristically monitoring the current size and complexity of the
  solver's formula.
\end{itemize}

One remaining function, \texttt{(ipasir-signature)}, is supported through another
mechanism, in case the user needs to access the solver library's version information.
The \texttt{ipasir-signature} function is
constrained to return a string.  When the backend is loaded, we define
a new function (in raw Lisp) that returns the result from the external
library's \texttt{ipasir_signature} call.  We use \texttt{defattach}
to attach this function to \texttt{ipasir-signature}.  Defattach is
designed to allow constrained functions to execute only in contexts
where their results can't be recorded as logical truths.  Otherwise we
could prove via functional instantiation that any function that always
returns a string returns the same value as \texttt{ipasir-signature}.
We still need to trust that the backend library always returns the
same string from \texttt{ipasir_signature}; otherwise we could prove
NIL using, for example, a clause processor that checks
 \texttt{(equal (ipasir-signature) (ipasir-signature))}.

\section{Interfacing with the External Library Backend}
\label{section:backend}
The interface with the external IPASIR implementation library is defined
in the book \texttt{ipasir-backend}.  This loads (in
raw Lisp) an external shared library specified by the environment
variable \texttt{IPASIR\_SHARED\_LIBRARY}, then redefines the
executable functions of the abstract stobj interface so that they call
into the external library.  We use the Common Lisp CFFI (Common
Foreign Function Interface) package available through Quicklisp to
load and call functions in the shared library.

The real underlying object on which the ipasir interface functions are
run is a vector containing the following seven pieces of data:
\begin{enumerate}
\setcounter{enumi}{-1}
\item the foreign pointer to the backend solver object used by the C API
\item the foreign pointer to the structure tracking the callback count and limit
\item the current solver status (\texttt{:undef},
  \texttt{:input}, \texttt{:unsat}, or \texttt{:sat})
\item a Boolean saying whether the current new clause is empty
\item a Boolean saying whether the current history is nonempty (that
  is, whether the solver has ever been initialized)
\item a list of literals tracking the current assumption
\item a list of literals which is the previous solved assumption if
  status is \texttt{:unsat}.
\end{enumerate}

To install the backend interface, we redefine all the executable
interface functions of the abstract stobj so that they run the
appropriate operations on this vector.
  These executable interface functions are defined in ACL2 as
operating on the concrete stobj \texttt{ipasir\$c} that was used to
define the abstract stobj \texttt{ipasir}.  (We discuss a soundness
problem that arises from this redefinition in Section
\ref{section:extralogical}.)

The backend functions are mostly simple wrappers around calls of the
appropriate foreign library functions on the backend solver object (slot 0 of the vector),
along with a small amount of bookkeeping to maintain the other fields of the vector.
The wrappers also transform input literals
from SATLINK to DIMACS format, and translate some query results
to make them more idiomatic in Lisp.  The initialization functions
also handle errors that might occur due to the external library not
being loaded, catching the raw Lisp error and producing an ACL2 hard
error instead.  The \texttt{ipasir-set-limit\$c} function sets up a
callback (created using CFFI's \texttt{defcallback}) that counts the
number of times it is called and returns 1 to end the SAT search if a
limit is reached.  To support this, \texttt{ipasir-init\$c} and
\texttt{ipasir-reinit\$c} reset the count to 0 and the limit to
\texttt{NIL}, and \texttt{ipasir-solve\$c} resets the count to 0
before beginning the SAT search.

\section{Soundness Assessment}
\label{section:soundness}

While we can't show our integration to be sound via formal proof
---in fact, we know of one soundness bug, discussed in Section \ref{section:extralogical}---
we hope that it can be accepted through a social process of discussing
and eliminating potential problems.  Similar to ACL2 itself, we hope
that over time it can be inspected and determined to be largely sound.
We prioritize soundness problems that might be
encountered by accident and yield undetected false results, rather
than those which would need to be exploited intentionally.

To start the assessment of the IPASIR integration's soundness, we
think it is useful to sort the problems that could cause
unsoundness into three main categories.  These distinctions are
blurry, but useful at least in describing what we think is the state
of the integration's soundness or lack thereof.  We will describe the
three categories and then devote a subsection to each category and our
efforts to avoid unsoundness of that kind.

First, we may have soundness problems due to mismatches between the
behavior of the external library and our assumptions about that
behavior.  These might be bugs in the library or simply invalid
assumptions on our part.  Out of the three categories, we believe
these are the most critical, since they are most likely to yield
undetected false results.

Second, we may have other logical problems not directly due to invalid
assumptions about the behavior of the library.  We focus here on the
problem of ostensible functions that may actually return different
values on the same inputs.

Third, we may have soundness problems due to incidental misfeatures of
our integration mechanisms, rather than due to the logical modeling of
the external library.  We know of one soundness bug in this area that
remains unfixed, though it would be implausible for it to be exploited
unintentionally.

\subsection{Validity of Assumed Behavior}

We first discuss potential problems due to mismatches between the
assumed and actual behavior of the external incremental SAT library
(along with the raw Lisp code interfacing with it).  Of course, we
first have to assume that the external SAT library is bug-free: if the
external solver's \texttt{ipasir_solve} routine produces a wrong
answer, then our interface is not sound. Beyond this, our main defense
against these problems is to carefully assess what we are assuming
about each interface function.  We model our correctness argument on
the ACL2 proof obligations necessary to admit an abstract stobj
\cite{EPTCS114.5}.  That is, it suffices to prove, for some
correlation relation between the logical model state and the
underlying implementation state:

\begin{itemize}
\item The logical model and the underlying implementation of the stobj
  creator function produce initial values satisfying the correlation.
\item For each updater, if its logical model and implementation are
  passed objects satisfying the correlation and the updater's guard,
  their respective results satisfy the correlation.
\item For each accessor, if its logical model and implementation are
  passed objects satisfying the correlation and guard, their results
  are equal.
\item For each interface function, the guard of the logical model
  implies the guard of the implementation.
\item The logical model of each updater preserves the well-formedness
  predicate.
\end{itemize}

The last two items are trivial in our case: our implementations are in raw
Lisp and have no guards, and the well-formedness requirement has
nothing to do with the implementation side and thus is already proved
when admitting the abstract stobj in the logical model.

The other three requirements depend on the correlation relation that
we maintain.  We can't state this in the ACL2 logic since it involves
the raw Lisp and external C library implementation, but we
nevertheless try to describe it precisely:

\begin{itemize}
\item The \texttt{formula} must be logically equivalent to the set of
  clauses stored in the solver object, which is field 0 of the
  implementation vector.  (The implementation solver may simplify the
  formula, so it may not be stored in the same form as in the model.)
\item The \texttt{assumption} must reflect the set of assumption
  literals in the solver, and must also equal field 5 of the
  implementation vector.
\item The \texttt{new-clause} field must reflect the clause under
  construction of the solver object, and must additionally be empty if
  and only if field 3 of the implementation vector is true.
\item The \texttt{status} must be equal to that recorded in field 2 of
  the implementation vector, and also correspond to the solver
  object's current state.
\item The \texttt{solution}, when in the \texttt{:unsat} state, must
  correspond to the solver's recorded unsatisfiable subset of the
  assumptions, and when in the \texttt{:sat} state, must correspond to
  the solver's recorded satisfying assignment.
\item The \texttt{solved-assumption}, when in the \texttt{:unsat}
  state, must equal field 6 of the
  implementation vector, which must also be the set of assumptions
  from the last call of \texttt{ipasir-solve}.
\item The \texttt{callback-count} must equal the count of callbacks
  stored in field 1 of the implementation vector, or 0 of that field
  is a null pointer.
\item The \texttt{history} must be nonempty if and only if field 4 of
  the implementation vector is true.
\end{itemize}

There are 20 functions (including \texttt{create-ipasir} and all
accessors/updaters) in the abstract stobj interface.  Of these, nine
are purely accessors, 10 are purely updaters, and one
(\texttt{ipasir-solve}) is both.  To fully argue the correctness of
all of these with respect to the correlation relation above, we'd need
to justify the correlations for each of the eight fields for each of
the updaters, and additionally argue that the model and implementation
of each accessor return equal values when the correlation holds.  For
most interface functions, this is a straightforward but tedious
argument.  However, \texttt{ipasir-solve} is a special case that
requires additional explanation.

Rather than a full definition, \texttt{ipasir-solve} has constraints
that do not fully specify what its result must be.  In fact, its
constraints are not sufficient to prove that it preserves the
correlation relation.  For example, suppose that on some inputs the
implementation of \texttt{ipasir-solve} produces a satisfiable result
and therefore ended in the \texttt{SAT} state, but the logical model
instead produces \texttt{:failed} and therefore ends in the
\texttt{:input} state.  This is consistent with the constraints for
\texttt{ipasir-solve}, which allow it to return \texttt{:failed} for
any input.  But the Lisp definition of \texttt{ipasir-solve} simply
translates the result from the implementation library into the ACL2
idiom, so this can't happen---as discussed in Section
\ref{section:logicalmodel}, the logical model of \texttt{ipasir-solve}
is given by its observed behavior, not by its constraints.  We do need
the constraints to be consistent with all such concrete executions,
which we argue, as with the other interface functions, by appealing to
the API description and knowledge of what a SAT solver is supposed to
do.

\subsection{Logical Consistency}
% \section{Avoiding Unsoundness Due to Nondeterminism}
\label{section:nondeterminism}

Aside from mismatches between our logical assumptions and
implementation realities, other possible soundness problems may occur
due to inconsistencies in the logical story.  In particular, we'll
discuss how we addressed nondeterminism, which might otherwise cause
ostensible functions to return different results on the same inputs,
leading to unsoundness.  We have already discussed the incompleteness
of the constraints on \texttt{ipasir-solve}, which is another example
of such a problem; this was a soundness bug in a previous version of
the library, which we solved by adding unknown constraints to
\texttt{ipasir-solve}.

Nothing in the IPASIR interface specification implies that the solver
library must be fully deterministic.  We therefore need to expect that
the solver may produce different results given the same inputs --
i.e. the interfaces to the solver are not actually functions.  This
could easily lead to unsoundness.  Specifically, if we could arrange
for some interface function to be called twice on inputs that are
provably equal and return different results, we could use this to prove \texttt{NIL}.
Since we can't control the results returned by the
underlying solver, we instead ensure that we can't run a vulnerable
interface function twice on provably equal inputs.  We therefore seek
to prevent:
\begin{itemize}
\item \textit{Coincidence}: creation of two distinct \texttt{ipasir} objects
  that are provably equal
\item \textit{Recurrence}: recreation of a solver state provably
  equal to a previous state after changing it in a way that might
  affect the answers returned from queries.
\end{itemize}

To prevent \textit{coincidence}, we require that a solver must always
be initialized for the first time by \texttt{ipasir-init}, which seeds
the solver's history field with an object taken from the ACL2 state's
oracle field, removing that object from the oracle.  The oracle is a
mechanism by which ACL2 models nondeterminism; it is simply an object
about which nothing is initially known, and which can only be accessed
by \texttt{read-acl2-oracle}, which returns the oracle's \texttt{car}
and replaces the oracle with its \texttt{cdr} \cite{acl2:doc}.)  This ensures that
we can't prove anything about what
an oracle read will produce until that read happens.  The object read
from the oracle in this case is written to the solver's history field,
which has no accessors in the abstract stobj interface, so we can't
determine after the fact what object was read from the oracle, either.
Additionally, there is no interface function that removes elements
from the history, so that object remains there permanently.  Since we
can't determine the value stored in that field, and since any two
solver objects will be seeded upon initialization with two independent
oracle reads, we can't prove them to be equal once they are
initialized.

To prevent \textit{recurrence}, every operation that changes the
external library's state is modeled as consing some additional object
onto the history.  There is no operation that clears the history or
removes any element from it.  Therefore a solver object can never be
made to go back to a previous state, because the length of its history
always increases.

The abstract stobj interface functions which are just accessors (that
is, they don't return a modified \texttt{ipasir} object) are assumed
to be read-only and not affect the state of the underlying solver;
therefore, they don't need to update the history.  Of these,
\texttt{ipasir-get-curr-stats}, \texttt{ipasir-val}, and \texttt{ipasir-failed}
 make library calls that might affect
the external solver object, but if any of these affected the solver state
in an observable way we would view it as a bug in the external
library.  Additionally, \texttt{ipasir-input} doesn't need to update
the history because it doesn't touch the external solver object.

\subsection{Integration Artifacts}
\label{section:extralogical}

A third class of soundness problems arise from factors that we view as
unrelated to the logical story of the IPASIR integration; they are
merely artifacts of the ACL2 mechanisms that we used or abused in
order to achieve the integration.  In particular, the
\texttt{defabsstobj} event is almost exactly what we need to implement
an external library interface like this one.  However, this requires
us to supply a fake implementation using a concrete stobj, in our case
\texttt{ipasir\$c}.  This is perhaps unnecessarily complicated and is
fertile ground for unsoundness.

For example, in order to install the implementations of the abstract
stobj interface functions, we redefine the executable versions of
those functions, which were originally defined as operations on the
concrete stobj \texttt{ipasir\$c}. A soundness problem arises here
because users could apply these functions after redefinition to the
\texttt{ipasir\$c} object or any stobj congruent to
\texttt{ipasir\$c}.  But the interface to \texttt{ipasir\$c} is not
restricted in the same way as the \texttt{ipasir} abstract stobj---in
fact, it has several low-level accessors, such as one that purports to
retrieve the \texttt{ipasir\$a} object that logically models the
solver's behavior.  The use of such an accessor could easily lead to
unsoundness.  For example, the following function can easily be shown
to always return \texttt{T} as its first return value, but its
execution (after loading the backend) returns \texttt{NIL}:

\begin{verbatim}
(define ipasir$c-contra (state)
  (with-local-stobj ipasir$c
    (mv-let (ans state ipasir$c)
      (b* (((mv ipasir$c state) (ipasir-init$c ipasir$c state))
           (solver (ipasir-get ipasir$c)))
        (mv (ipasir$a-p solver) state ipasir$c))
      (mv ans state))))
\end{verbatim} %$

To prevent this, we make the redefined functions \textit{untouchable},
which disallows users from calling these functions or defining new
functions that call them.  Unfortunately, this also prevents the
creation of new abstract stobjs congruent to \texttt{ipasir}, which is
often desirable.  Also unfortunately, this mitigation doesn't
completely solve the problem.  A determined user can defeat the
untouchability of any function defined outside the ACL2 system as
follows: copy all the events needed to admit the function, define a
wrapper for that function, then load the book that declares the
function untouchable.  After that point, simply call the wrapper
instead of the untouchable function.

We hope to solve this problem more comprehensively in future work,
perhaps by adding some features to the ACL2 system.  A relatively easy
solution for the specific problem above would be to allow concrete
stobjs to be defined with non-executable accessors and updaters.  Then
the only functions that could be executed on that stobj would be ones
that were redefined under the hood---namely, the abstract stobj
interface functions.  A more heavyweight but perhaps also more direct
solution would be to extend ACL2 with support for a
\texttt{defabsstobj} variant intended for this sort of application,
perhaps avoiding the introduction of an underlying concrete stobj
altogether.

One other known extralogical problem has to do with ACL2's
\texttt{save-exec} feature.  Under normal circumstances this feature
can be used to save an executable memory image of the running ACL2, so
that it can be restarted from the current state.  However, foreign
objects cannot be saved in the heap image.  Therefore, if we save an
executable in which the global \texttt{ipasir} object is initialized,
the underlying solver object will not exist
when the image is executed.  Running any ipasir interface functions
then will at minimum cause a raw Lisp error and could potentially
cause memory corruption or unsoundness.  This problem can be avoided
by ensuring that live ipasir stobjs are in the \texttt{:undef} state
before saving an executable.

\section{Application: AIG SAT sweeping}
\label{section:fraiging}

We built on the IPASIR integration to implement SAT sweeping, or
fraiging, on top of the AIGNET and-inverter graph (AIG) library
\cite{EPTCS114.8}.  Circuit structures such as AIGs are often a good
target for incremental SAT, since the logical relationships among the wires can be encoded
in the permanent formula and the various queries encoded in the
assumptions.  During each SAT check, the solver accumulates learned
clauses and heuristic information about the circuit that can be used
on subsequent checks.%   After some point the solver's clause database
% becomes too large and burdened with information about parts of the
% circuit that are no longer in active use, and it is beneficial to then
% start over with a new solver object.

The purpose of SAT sweeping \cite{Mishchenko:2006:ICE:1233501.1233679}
is to search for and remove redundancies in the AIG; that is, to find
pairs of nodes that are provably equivalent and remove one of them,
connecting its fanouts to the other.  This reduces the size of the
graph and speeds up subsequent algorithms while preserving
combinational equivalence; that is, the combinational formulas of
corresponding outputs or next-states in the input and output networks
are equivalent.  This is a powerful algorithm for combinational
equivalence checking because often the two circuits contain many
equivalent internal nodes, and finding these equivalent internal nodes
makes it much easier to prove the full circuits equivalent.

\newcommand{\cnfvar}[1]{\ensuremath{\mathrm{cnf}(#1)}}

As a preliminary requirement for SAT sweeping, we need to be able to
use SAT to check equivalences between AIG nodes.  We use a standard
Tseitin transformation \cite{Tseitin1983} to encode substructures of
the AIG into CNF as needed.  Whenever we need to do a SAT check
involving some node, we encode the fanin cone of that node into the
CNF formula.  This results in a CNF variable corresponding to that
node.  This process maintains the invariant that each evaluation of
the AIG maps to a satisfying assignment of the CNF formula, where for
each AIG node that has a corresponding CNF variable, the assignment to
the variable is the same as the value of the node:
\begin{verbatim}
(forall (invals regvals cnf-vals)
        (equal (satlink::eval-formula
                (ipasir::ipasir$a->formula ipasir)
                (aignet->cnf-vals
                 invals regvals cnf-vals sat-lits aignet))
               1))
\end{verbatim}%$
In the above formula, \texttt{sat-lits} is a stobj containing a
bidirectional mapping between AIG literals and SAT literals.  The
\texttt{invals} and \texttt{regvals} are assignments to the AIG's
primary inputs and registers (which are treated as combinational
inputs for this purpose).  The function \texttt{aignet->cnf-vals} maps
the evaluation of the AIG given by \texttt{invals} and
\texttt{regvals} to an assignment of the CNF variables, replacing the
relevant slots of stobj \texttt{cnf-vals}; specifically, it satisfies:
\begin{verbatim}
(implies (sat-varp m sat-lits)
         (equal (nth m (aignet->cnf-vals
                         invals regvals cnf-vals sat-lits aignet))
                (lit-eval
                 (sat-var->aignet-lit m sat-lits)
                 invals regvals aignet)))
\end{verbatim}
That is, each variable in the CNF formula is assigned the evaluation
of its corresponding AIG literal, namely \texttt{(sat-var->aignet-lit
  m sat-lits)}.  The invariant above says that this is always a
satisfying assignment for the CNF formula.  Therefore, if we obtain an
UNSAT result, it must be that the added assumptions,
\texttt{(ipasir\$a::assumption ipasir)}, are to blame. In particular,
an UNSAT result implies that no evaluation of the AIG yields a CNF variable assignment under which
the assumption literals are all simultaneously true; therefore, the
corresponding AIG literals can't be simultaneously true either.  So
to check the
equivalence of AIG nodes $a$ and $b$, we can do two SAT checks
after encoding both nodes into CNF: one with assumption $\cnfvar{a}
\wedge \neg \cnfvar{b}$, and one with assumption $\neg \cnfvar{a}
\wedge \cnfvar{b}$.  If both these checks return UNSAT, we can then
conclude that there is no evaluation of the AIG in which the values of
nodes $a$ and $b$ differ.

To perform SAT sweeping, we begin with a set of potential equivalences between the
nodes, derived by random simulation.  We then sweep through the graph
in topological order, meaning all of a node's fanins must be processed
before we process that node.  As we sweep, we build a copy of the
graph with redundancies removed, and maintain a mapping from the
processed nodes of the input graph to their combinationally-equivalent analogues in the output
graph.  To sweep a node $Q$, we first create a new node $Q'$ in the
output graph whose fanins are the analogues of the fanins of $Q$.  If
$Q$ has no potential equivalences or all potential equivalences occur
later in the topological order (and therefore haven't yet been
processed), we set $Q'$ as the analogue of $Q$ and continue with the
next node.  Otherwise let $P$ be the potentially equivalent node
earliest in the topological ordering and let $P'$ be its analogue in
the output graph.  We check using SAT whether $Q'$ and $P'$ are
equivalent.  We set the analogue of $Q$ to $P'$ if they are equivalent
and $Q'$ if not.  If the SAT check produces a counterexample (rather
than failing due to a solver limit), we simulate the circuit using
that counterexample and refine the candidate equivalence classes to
account for any newly differentiated pairs of nodes.

We have proved in ACL2 that this algorithm produces a new AIG that is
combinationally equivalent to the input AIG.  The correctness proof is
based on the invariant that the mapping of nodes of the input graph to
their analogues in the output graph preserves combinational
equivalence.  At a given step, we set the mapping for $Q$ either to
$Q'$, which is equivalent to $Q$ because it is the same operator
applied to fanins which are equivalent by inductive assumption to the
fanins of $Q$, or to $P'$ in the case where $P'$ has been shown by SAT
to be equivalent to $Q'$.

\section{Conclusion}

This integration of incremental SAT solvers via the IPASIR API is in
everyday use for hardware verification at Centaur, largely through the
fraiging transform described above.  The library and the fraiging
algorithm are both available in the ACL2 community books, in
directories \texttt{centaur/ipasir} and \texttt{centaur/aignet},
respectively.

The soundness of this integration is a work in progress.  One
soundness bug is known to exist, though it doesn't pose a practical
risk of undetected false results.  We hope to address this problem in
future work, though that might require changes to ACL2 itself.  Other
soundness problems may be revealed with further study, but we hope
that the basic approach has the potential to be sound.

\bibliographystyle{eptcs}
\bibliography{../bib}
\end{document}